\documentclass[a4paper]{article}

\usepackage{INTERSPEECH2020}
\usepackage{multirow}
\usepackage{bm}
\usepackage{cite}

\title{SAN-M: Memory Equipped Self-Attention for End-to-End Speech Recognition}
\name{Zhifu Gao$^1$, Shiliang Zhang$^1$,  Ming Lei$^1$, Ian McLoughlin$^2$}
\address{
	$^1$Speech Lab, Alibaba DAMO Academy \\
	$^2$ICT Cluster, Singapore Institute of Technology
}
\email{\{zhifu.gzf, sly.zsl, lm86501\}@alibaba-inc.com, ian.mcloughlin@singaporetech.edu.sg}

\begin{document}

\maketitle
\begin{abstract}

End-to-end speech recognition has become popular in recent years, since it can integrate the acoustic, pronunciation and language models into a single neural network.
Among end-to-end approaches, attention-based methods have emerged as being superior.
For example, \emph{Transformer}, which adopts an encoder-decoder architecture.
The key improvement introduced by Transformer is the utilization of self-attention instead of recurrent mechanisms, enabling both encoder and decoder to capture long-range dependencies with lower computational complexity.
\
In this work, we propose boosting the self-attention ability with a DFSMN memory block, forming the proposed memory equipped self-attention (SAN-M) mechanism.
Theoretical and empirical comparisons have been made to demonstrate the relevancy and complementarity between self-attention and the DFSMN memory block. Furthermore, the proposed SAN-M provides an efficient mechanism to integrate these two modules. 
\
We have evaluated our approach on the public AISHELL-1 benchmark and an industrial-level 20,000-hour Mandarin speech recognition task. On both tasks, SAN-M systems achieved much better performance than the self-attention based \emph{Transformer} baseline system. Specially, it can achieve a CER of 6.46\% on the AISHELL-1 task even without using any external LM, comfortably outperforming other state-of-the-art systems.
\end{abstract}


\section{Introduction}
\label{sec:intro}

Conventional automatic speech recognition (ASR) systems usually adopt the hybrid architecture \cite{dahl2011context}, which consists of separate acoustic, pronunciation and language models (AM, PM, LM).
\
Recently, so-called \emph{end-to-end}~(E2E) approaches have rapidly gained prominence in the speech recognition community.
End-to-end ASR systems fold the AM, PM and LM into a single neural network that dramatically simplifies the training and decoding pipelines.
Two popular approaches for this are neural networks with Connectionist Temporal Classification (CTC) -like criteria \cite{graves2006connectionist, graves2013speech} and attention-based models~\cite{chan2016listen,kim2017joint}.
\
The CTC-based approach has demonstrated its superiority over hybrid architecture, however, it requires an external LM for good performance~\cite{graves2014towards,sak2015fast}.
\
Unlike CTC-based approaches, attention-based models generate character sequences without any unreasonable independence assumption between characters, which enables it to effectively learn an implicit language model.

A typical attention-based model could be divided into two main components; an encoder and a decoder, which are jointly trained towards maximizing the likelihood of target sequences generated from acoustic feature sequences.
\
In early works~\cite{chan2016listen,bahdanau2016end}, long short-term memory neural networks (LSTMs) were widely used to model long-term dependencies among acoustic features in the encoder and output sequences in the decoder.
The attention module inside the decoder interacts between the output representations of the encoder and the hidden states of the decoder, to compute context vectors.
\
LSTM-type networks have a strong ability to capture long-term dependencies within the sequential data using the mechanism of recurrent feedback.
However, they suffer from the high computational complexity and a `painful' training process, \emph{i.e.}, gradient vanishing~\cite{bengio1994learning}.
Therefore, many authors have been inspired to search for more computationally-efficient and flexible architectures for sequential modeling.
\begin{figure}[t]
	\centering
	\includegraphics[width=0.9\linewidth]{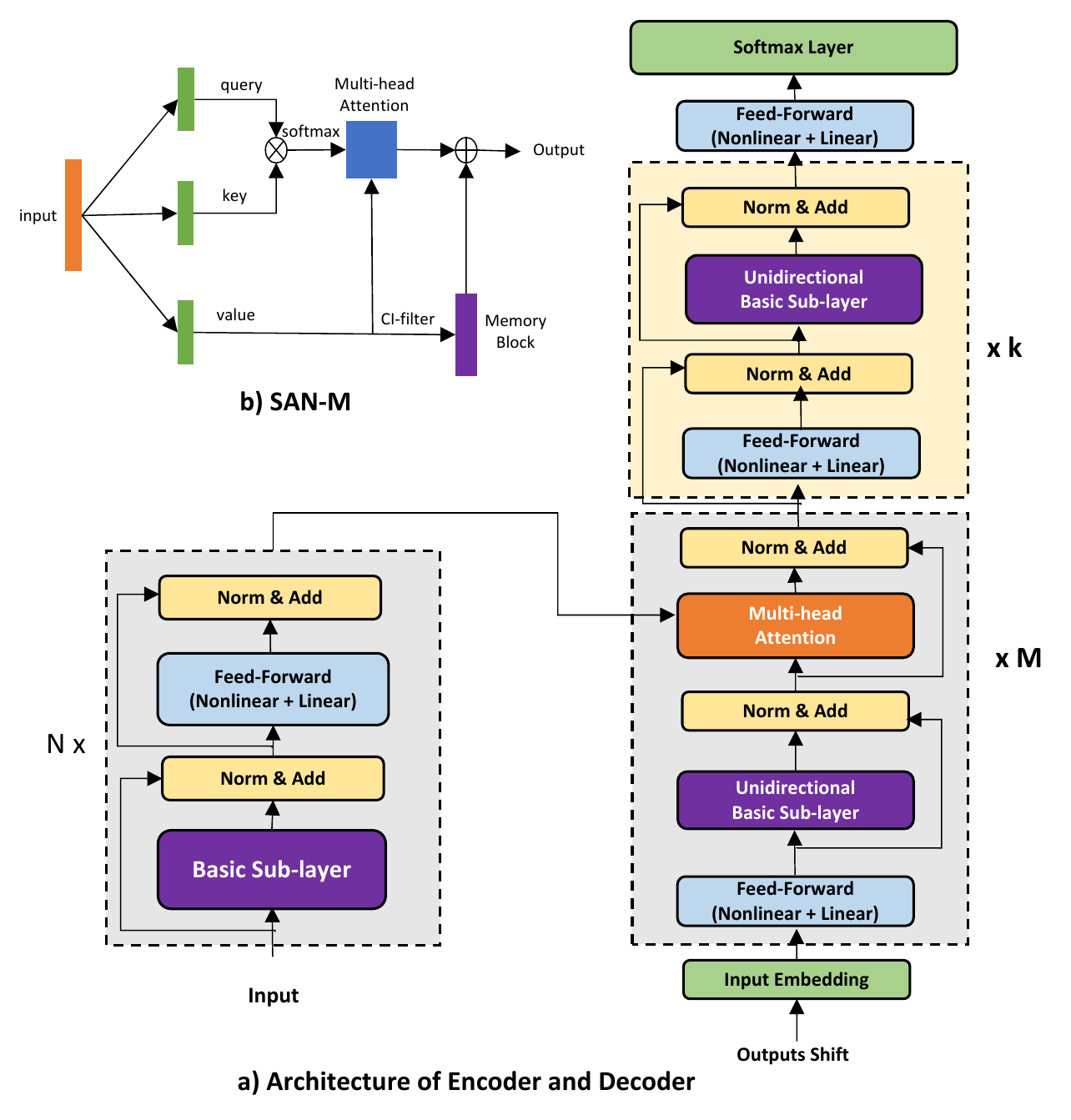}
	\caption{Illustration of: a) the architecture of encoder and decoder. b) the SAN-M architecture (top left).}
	\label{fig:dfsmn}
	\vspace{-3mm}
\end{figure}

\
In the past few years, some efficient models, \emph{e.g.}, convolutional neural networks~\cite{abdel2014convolutional} and time-delay neural networks~\cite{peddinti2015time}, have been employed to improve the training process.
Specially, Zhang $et~al.$ proposed a deep feed-forward sequential memory network (DFSMN) to replace LSTM in hybrid architectures~\cite{zhang2015feedforward, Zhang2018Deep} and in CTC-based models~\cite{zhang2018acoustic,zhang2019investigation}.
\
More recently, Transformer has become popular in seq2seq tasks, \emph{e.g.}, neural machine translation~\cite{vaswani2017attention}, ASR~\cite{pham2019very,dong2018speech,tian2019self,zhang2020transformer}, and has shown very promising performance.
\
The key improvement is the utilization of self-attention instead of recurrent models, \emph{e.g.}, LSTM, 
to model feature sequences in both encoder and decoder. 
This enhances the ability to capture long-range dependencies with lower computational complexity and to enable more parallelizable training.
\

Both self-attention and DFSMN memory blocks was proposed to replace LSTM for sequential modeling.
Self-attention has powerful long-term dependency modeling abilities inside the full sequence~\cite{vaswani2017attention}.
Unlike self-attention, a single DFSMN memory block layer was designed to model local-term dependencies, with the long-term contexts captured by stacking multiple layers~\cite{zhang2015feedforward}. 
\
To some extent, the self-attention and DFSMN memory block seems complementary for each other. 
Thus, in this work, we aim to design a new structure that exploits the complementarity between self-attention and DFSMN memory blocks, under the framework of attention-based models.
\
Firstly, we have made theoretical and empirical comparisons between self-attention and DFSMN memory blocks.
\
Secondly, we have designed a new structure called memory equipped self-attention~(SAN-M) to effectively combine the strength of both.
You $et~al.$ proposed inserting self-attention layers into DFSMN for hybrid architectures~\cite{you2019dfsmn}.
By contrast, we propose incorporating these into E2E ASR models.
Furthermore, the proposed SAN-M combines them both within a single basic sub-layer, in deep-fusion style.
\

We report extensive experiments on the public AISHELL-1 benchmark and an industrial-level 20,000-hour Mandarin speech recognition task. On both tasks, SAN-M based systems achieve much better performance than the self-attention based \emph{Transformer} baseline system. Specially, achieving 6.46\% CER on AISHELL-1 even without an external LM, which is the best performance on this task to date (shown later in Table.\ref{tab:AISHLL_state_of_art}).

\section{The proposed methods}                       
\label{sec:DFSMN-Transformer}
\begin{figure}[t]
	\centering
	\includegraphics[width=0.9\linewidth,height=5cm]{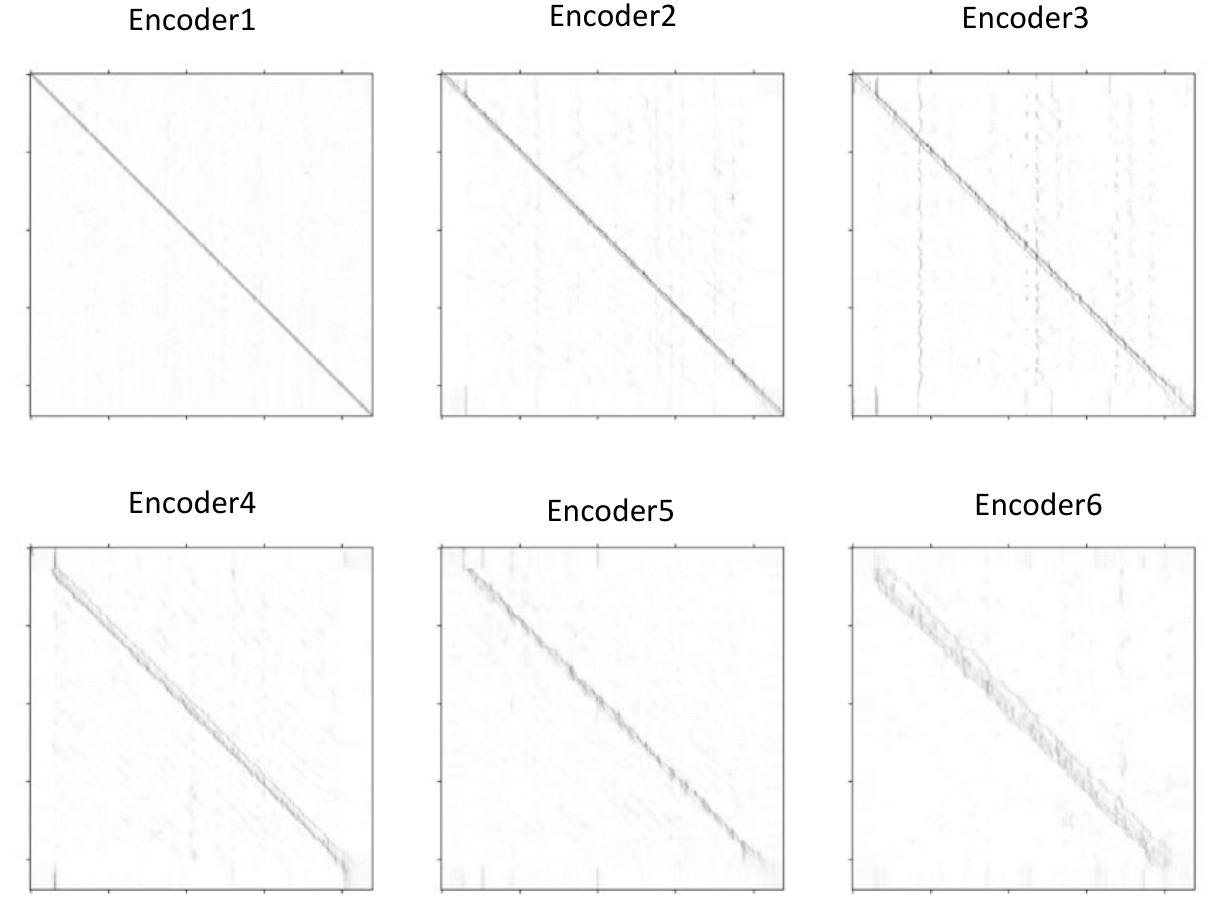}
	\caption{Self-attention image maps from different encoder layers for a given sequence.}
	\label{fig:transformer_encoder}
\end{figure}

\subsection{Overview}
\label{overview}
Transformer was first proposed for neural machine translation~\cite{vaswani2017attention}, where it obtained state-of-the-art results on many tasks.
It was then introduced into speech processing tasks, $e.g.$, ASR~\cite{dong2018speech,pham2019very} and text-to-speech~\cite{lakew2018comparison}.

As shown in Fig.~\ref{fig:dfsmn} a), our network follows the overall architecture of  Transformer~\cite{vaswani2017attention}, which consists of an encoder and a decoder.
\
The former maps an input sequence $\mathbf{X}$ to a sequence of hidden representations $\mathbf{Z}$ and consists of $N$ blocks of basic sub-layer and feed-forward sub-layer.
\
The decoder, meanwhile, generates one element of output sequence $\mathbf{Y}$ at each time step, consuming representations $\mathbf{Z}$.
\
As an auto-regressive decoder, it consumes the previously produced characters as additional inputs when producing the next character at each step~\cite{graves2013generating}.
\
It consists of three components. 
The first components is $M$ blocks which each consist of a feed-forward sub-layer, a unidirectional basic sub-layer and a multi-head attention sub-layer.
Then $K$ blocks which each comprise a feed-forward and a unidirectional basic sub-layer.
The last component is a single feed-forward sub-layer to output characters.

In this paper, we will firstly give a brief review of self-attention and DFSMN memory block.
Then we will incorporate them into the (unidirectional) basic sub-layer respectively. 
Theoretical analysis and empirical result comparisons will determine the strength of this approach.
Furthermore, we present the proposed new structure, memory equipped self-attention~(SAN-M) as the (unidirectional) basic sub-layer to effectively combine the strength of self-attention and DFSMN.

\begin{figure}[t]
	\centering
	\includegraphics[width=0.9\linewidth,height=3cm]{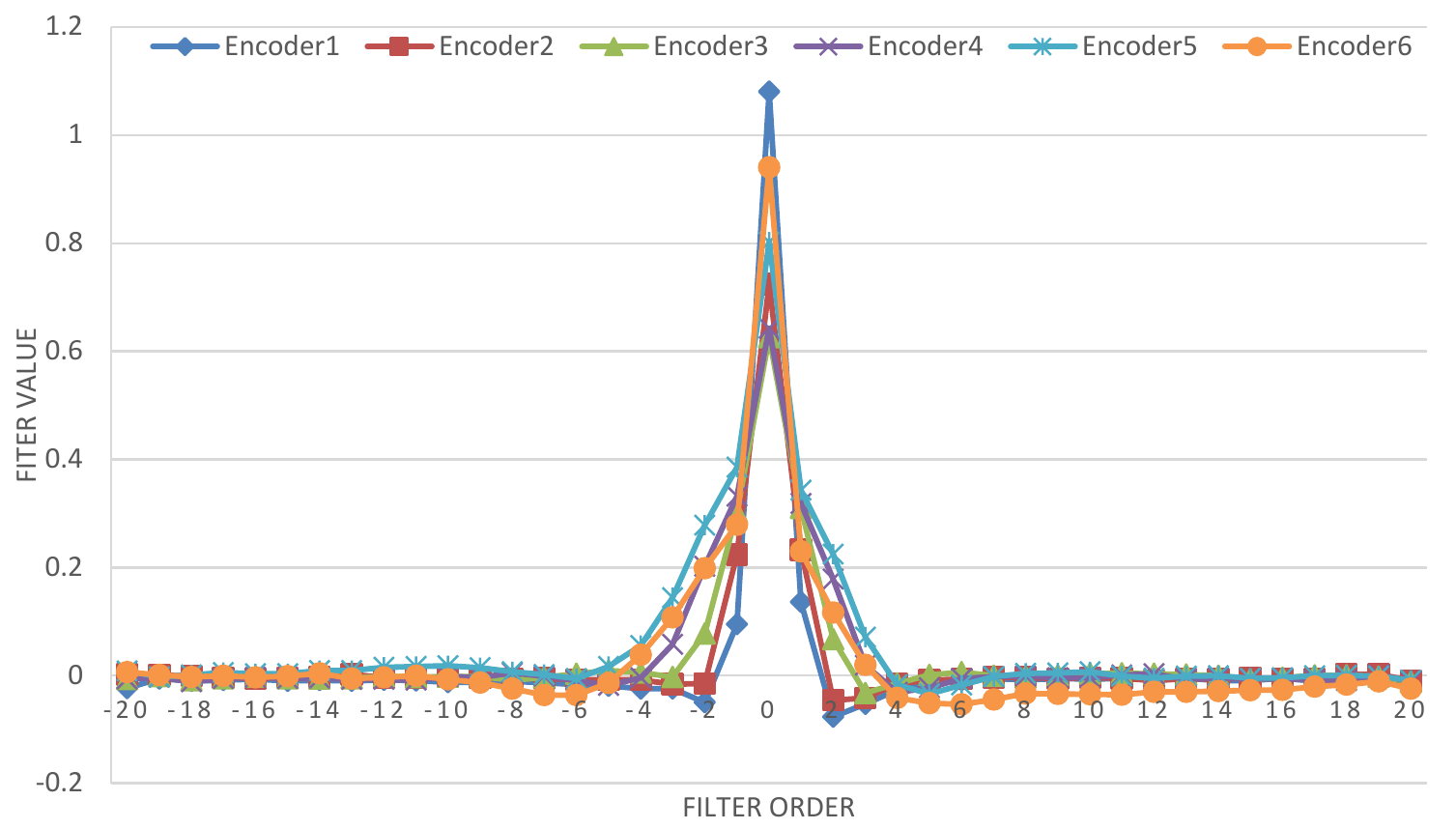}
	\caption{Visualization of the learned average filters of DFSMN memory blocks in different encoder layers.}
	\label{fig:dfsmn_encoder}
\end{figure}
\subsection{Multi-Head Attention}
Multi-head attention was proposed to jointly attend information from different representation subspaces at different positions~\cite{vaswani2017attention}.
It could be formulated as:
\begin{equation}\label{eq.multihead_self_attention}
{\rm MultiHead}(\mathbf{\mathbf{Q}, \mathbf{K}, \mathbf{V}}) = [{\rm head_{1}, ..., head_{h}}]\mathbf{W}^{O}
\end{equation}
\begin{equation}\label{eq.multihead_self_attention_i}
{\rm head_{i}} = {\rm Attention}(\mathbf{Q}_{i}, \mathbf{K}_{i}, \mathbf{V}_{i})
\end{equation}
\begin{equation}\label{eq.multihead_qkv}
(\mathbf{Q}_{i}, \mathbf{K}_{i}, \mathbf{V}_{i}) = (\mathbf{HW}_{i}^{Q}, \mathbf{XW}_{i}^{K}, \mathbf{XW}_{i}^{V})
\end{equation}
Where $\mathbf{Q}$, $\mathbf{K}$, $\mathbf{V}$ are queries, keys and values respectively. The projections are parameter matrices $\mathbf{W}_{i}^{Q}\in \mathbb{R}^{d_{model\times d_k}}$, $\mathbf{W}_{i}^{K}\in \mathbb{R}^{d_{model\times d_k}}$, $\mathbf{W}_{i}^{V}\in \mathbb{R}^{d_{model\times d_v}}$ and $\mathbf{W}^{O}\in \mathbb{R}^{hd_{v}\times d_{model}}$. $h$ is the number of heads, $d_{model}$ is the model dimension and $d_{k}$ is the key dimension.
$\mathbf{X} \in \mathbb{R}^{T \times d_{model}}$ and $\mathbf{H} \in \mathbb{R}^{T' \times d_{model}}$ are the inputs.
\
For each head, ``scaled dot-product attention''~\cite{vaswani2017attention} was adopted as the attention mechanism.
\
Given that, the outputs are formulated as:
\begin{equation}\label{eq.attention}
{\rm Attention}(\mathbf{Q}_{i}, \mathbf{K}_{i}, \mathbf{V}_{i}) = {\rm softmax}\left\{\dfrac{\mathbf{Q}_{i}\mathbf{K}_{i}^{T}}{\sqrt{d_{k}}}\right\}\mathbf{V}_{i}
\end{equation}

\subsection{Memory Block}

DFSMN~\cite{Zhang2018Deep} improved on the FSMN architecture by introducing skip connections and memory strides. 
It consists of three components: a linear projection, a memory unit and a weight connection from memory unit to the next hidden sub-layer.
The key elements in DFSMN are the learnable FIR-like memory blocks, which are used to encode long-context information into a fixed-size representation. As a result, DFSMN is able to model long-term dependencies in sequential data without using recurrent feedback. The operation in the $l$-th memory block takes the following form:
\
\begin{equation}
{\bf h}_t^\ell  = \max ({{\bf{W}}^\ell }{\bf m}_t^{\ell  - 1} + {\bf b}_t^\ell , 0)
\end{equation}
\begin{equation}
{\bf p}_t^\ell  = {\bf{V}}_t^\ell {\bf h}_t^\ell  + {\bf v}_t^\ell 
\end{equation}
\begin{equation}\label{eq.DFSMN}
{\bf m}_t^{\ell}={\bf m}_t^{\ell  - 1}+ {\bf p}^\ell_t+\sum\limits_{i = 0}^{N^\ell_1} {\bf a}_i^{\ell}  \odot {\bf p}^{\ell}_{t - s_1*i} + \sum\limits_{j = 1}^{N^\ell_2} {\bf c}_j^{\ell} \odot {\bf p}^{\ell}_{t + s_2*j}
\end{equation}
\begin{equation}\label{eq.M}
{\bf M}^{\ell} = [{\bf m}_1^{\ell}, {\bf m}_2^{\ell}, ..., {\bf m}_T^{\ell}]
\end{equation}
Here, ${\bf M}^{\ell}$ is the memory block. ${\bf h}_t^\ell$  and ${\bf p}_t^\ell$ denote the outputs of the ReLU layer and linear projection layer respectively. ${\bf m}_t^{\ell}$  denotes the output of the $\ell$-th memory block. $N^\ell_1$ and $N^\ell_2$ denote the look-back and lookahead order of the $\ell$-th memory block, respectively, while $s_1$ and $s_2$ are their respective stride factors.

\subsection{Comparing Self-Attention and Memory Blocks}
\label{comparison}

\begin{figure}[t]
	\centering
	\includegraphics[width=0.9\linewidth, height=2.3cm]{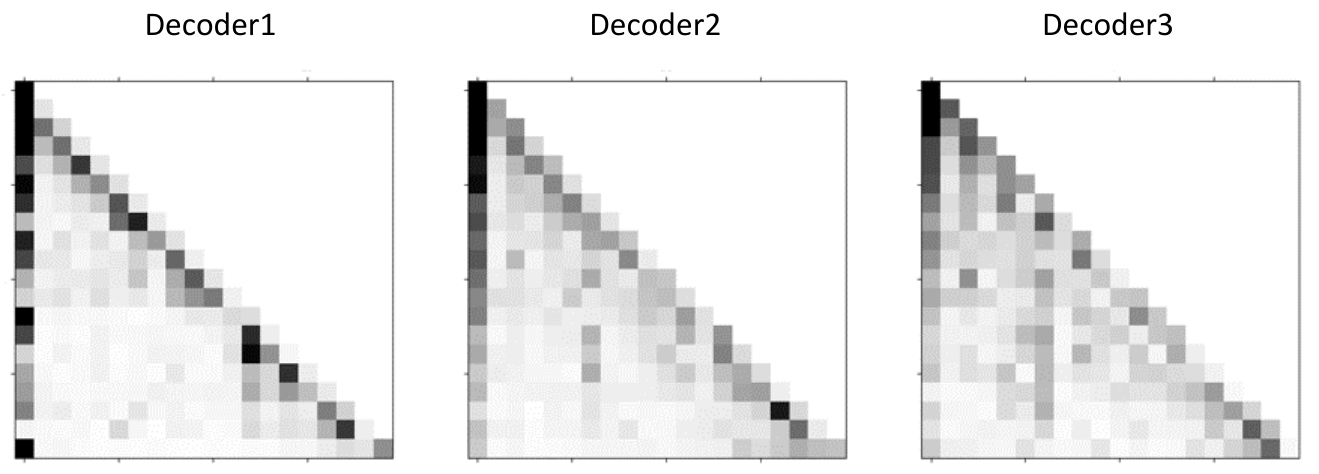}
	\caption{Image maps from three decoder layers to illustrate self-attention from a given sequence.}
	\label{fig:transformer_decoder}
\end{figure}
\begin{figure}[t]
	\centering
	\includegraphics[width=0.9\linewidth, height=3cm]{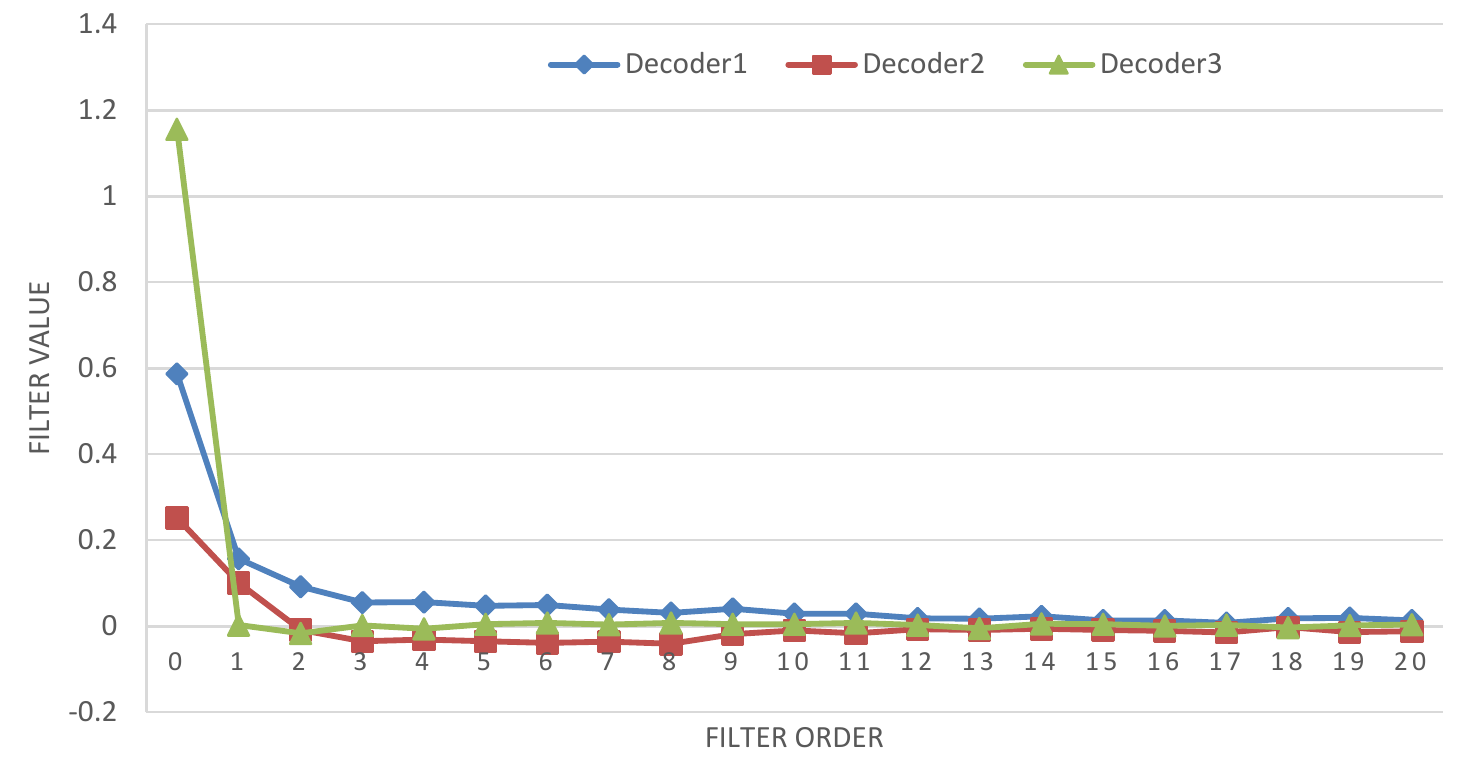}
	\caption{Visualization of the learned average filters in DFSMN memory blocks from different decoder layers(mirrored).}
	\label{fig:dfsmn_decoder}
\end{figure}
In this section, we will make an in-depth comparison between self-attention and DFSMN memory blocks.
\
Self-attention is an attention mechanism where the $queries$, $keys$ and $values$ are from the same sequence in Eq.~(\ref{eq.attention}).
Then the attention vector $\mathbf{c}_t$ is calculated as:
\
\begin{equation}\label{eq.att_vector}
\mathbf{c}_t = \sum_{j=0}^{T}{\alpha_{t,j}\mathbf{h}_j} =\
\sum_{i=0}^{t} {\alpha_{t,i}\mathbf{h}_i} + \sum_{j=t+1}^{T} {\alpha_{t,j}\mathbf{h}_j}
\end{equation}
\
Where $\bm{\alpha}_{t}=(\alpha_{t,1},...,\alpha_{t,T})$ are the weights of self-attention from a sequence at time $t$.
\
In terms of formulation, Eq.~(\ref{eq.att_vector}) is similar to the scalar FSMN memory block proposed in~\cite{zhang2015feedforward}.
If we take multi-head attention into consideration, Eq.~(\ref{eq.multihead_self_attention}) is similar to the vectorized FSMN memory block defined in Eq.~(\ref{eq.M}).
\
To summarise, the outputs of both DFSMN memory block and self-attention are computed by weighting and then summing the feature vectors.
The important difference is how to derive the weights.
\

As for self-attention, weights are calculated dynamically depending on the features themselves, which could be viewed as context-dependent (CD) coefficients.
This could learn time dependencies inside the full sequence. However, it may not be efficient since it must compute every time pair in the full sequence. The computational complexity is thus $\bm{O}(n^{2} \cdot d)$ \footnote{$n$ and $d$ are the length and dimension of a sequence respectively.}.
\

In terms of DFSMN memory block, weights are context-independent (CI) coefficients, and we could view this as learning the statistical average distribution of the whole dataset.
As defined in Eq.~(\ref{eq.DFSMN}), the range of context dependencies is controlled by $N^\ell_1$ and $N^\ell_2$, which means it is more computationally efficient and flexible.
The computational complexity is $\bm{O}((N^\ell_1 + N^\ell_2) \cdot n \cdot d)$.
Though the receptive field of a single layer is small, it can still model long-range dependencies by stacking multiple layers. 
\

We plot the CD-coefficients of self-attention in different encoder layers for a given sequence in Fig.~\ref{fig:transformer_encoder}.
A strong diagonal component is evident, which grows more diffuse and wider as we progress through deeper layers.
\
This reveals that learned features are mainly locally dependent, even though self-attention is able to model long-term dependencies over the full sequence.
\
In Fig.~\ref{fig:dfsmn_encoder}, showing the CI-coefficients of DFSMN memory for the same encoder blocks, we see a shape resembling a tower whose width increases as we progress through deeper layers.
\
For comparison, we also plot the self-attention matrix weights and DFSMN memory block vectors for the same decoder layers in Figs.~\ref{fig:transformer_decoder} and \ref{fig:dfsmn_decoder}.
These show that self-attention has learned much longer-range dependencies than the DFSMN memory block.
Our investigations have found that in practice, self attention for acoustic features in the encoder is often dominated by short-term dependencies. Consequently, it may therefore not be effectively capturing longer-term dependencies.

From the discussion above, we can briefly summarise:
(a) Self-attention has the ability to learn long-range dependencies inside the full sequence, 
yet the learned features are not necessarily always long-term dependent, particularly in the encoder.
\
(b) DFSMN memory blocks tend to learn local dependencies. Meanwhile they are more computationally efficient, and flexible, than self-attention.
\
(c) While self-attention learns long-term context dependencies focusing on single features, DFSMN memory blocks learn local-term dependencies from the statistical average distribution over the whole dataset,  meaning that they may well be more robust in practice.
\
\newcommand{\pp}[1]{\raisebox{-1.2ex}[0pt][0pt]{\shortstack{#1}}}
\begin{table}[t]
	\centering
	\caption{Performance comparison of three basic sub-layer types on AISHELL-1.}
	\begin{tabular}[t]{|c|c|c|c|c|}
		\hline
		\multirow{2}*{Encoder} & \multirow{2}*{Decoder} & \pp{Parameter(M)} & \multicolumn{2}{|c|}{CER(\%)} \\\cline{4-5}
		&       &         &   Dev & Test \\\hline \hline
		SAN & SAN & 46 & 6.58 & 7.33 \\\hline
		DFSMN & DFSMN & 37 & 5.92 & 6.81 \\\hline
		SAN-M & DFSMN & 43 & 5.74 & 6.46 \\\hline
	\end{tabular}
	\label{tab:AISHLL_res1}
\end{table}
\begin{table}[t]
	\centering
	\caption{State-of-the-art comparison on AISHELL-1.}
	\begin{tabular}[t]{|c|c|c|c|c|}
		\hline
		\pp{Model} & \pp{E2E} & \pp{LM} & \multicolumn{2}{|c|}{CER(\%)} \\\cline{4-5}
		&          &         &   Dev & Test \\\hline \hline 
		TDNN-LFMMI~\cite{bu2017aishell}  &    N     &   Y     &    6.44  &  7.62 \\\hline
		SA-T~\cite{tian2019self}   &    Y     &   N     &  8.30 &  9.30 \\\hline
		LAS~\cite{shan2019component} & Y      &   Y     &  -    & 8.71 \\\hline
		Joint CTC/attention~\cite{karita2019comparative} & Y & Y  & 6.00 & {6.70} \\\hline
		Proposed \bf{SAN-M} & Y  & N & \bf{5.74} & \bf{6.46} \\\hline         
	\end{tabular}
	\label{tab:AISHLL_state_of_art}
\end{table}

\begin{table*}[t]
	\centering
	\caption{Comparison of models on the 20000-hour Mandarin speech recognition task.}
	\vspace{-2mm}
	\begin{tabular}[t]{|c|c|c|c|c|c|c|c|}
		\hline
		Models &	CTC1 &	CTC2 &	EXP1 &	EXP2 &	EXP3 &	EXP4 &	EXP5 \\\hline\hline
		Encoder &	DFSMN &	DFSMN &	SAN &	DFSMN &	SAN-M &	SAN-M &	SAN-M \\\hline
		Decoder &	- &	- &	SAN &	DFSMN &	DFSMN &	DFSMN &	DFSMN \\\hline
		$d_{basic}$ - $d_{ffn}$ &	- &	- &	512-2048 &	512-2048 &	512-2048 &	256-1024 &	320-1280 \\\hline
		N &	- &	- &	10 &	10 &	10 &	40 &	40 \\\hline
		M &	- &	- &	6 &	6 &	6 &	6 &	6 \\\hline
		K &	- &	- &	0 &	0 &	0 &	6 &	6 \\\hline		
		Parameter (M) &	25 &	45 &	59 &	47 &	55	& 42	& 63 \\\hline
		Common Set (CER\%) &	11.6 &	9.9 &	9.8 &	10.2 &	9.4 & 	9.0	& 8.3	\\\hline
		Far-field Set (CER\%) &	20.3 &	17.7 &	15.0 &	16.7 &	14.3 &	13.7	& 12.5 \\\hline		
		
	\end{tabular}
	\label{tab:fly2w_res}
\end{table*}

\subsection{Memory Equipped Self-Attention}

From Section~\ref{comparison}, we found that self-attention tends to learn CD-dependencies within a single feature whereas DFSMN memory blocks tend to learn CI-dependencies from the statistical average distribution of whole dataset.
We think that the two structures might therefore be complementary to each other.
Following that insight, we designed memory equipped self-attention (SAN-M) to combine the strengths of both approaches.
As shown in Fig.~\ref{fig:dfsmn} b), a DFSMN filter has been added on the $values$ inside the $Multi$-$Head~Attention$ to output memory block.
The memory content is then added to the output of the $Multi$-$Head~Attention$, which could be formulated as:
\begin{equation}\label{eq.san-m}
\mathbf{Y} = {\rm MultiHead}(\mathbf{Q}, \mathbf{K}, \mathbf{V}) + \mathbf{M(V)}
\end{equation}
Where $\mathbf{Y}$ denotes the output of SAN-M. Unidirectional SAN-M means that both self-attention and DFSMN memory blocks themselves are unidirectional.

\section{Experiments}
\label{sec:exp}
\subsection{Experimental Setup}

We conduct extensive experiments to evaluate the performance of self-attention, DFSMN memory block and the combined SAN-M on Mandarin speech recognition tasks. 
We report results on the 170-hour AISHELL-1 released in~\cite{bu2017aishell} and an industrial-level 20000-hour-task described in~\cite{zhang2019investigation}, collected from multiple domains including news, sport, tourism, game, literature, education etc. It is divided into a training set and a development set in the ratio of 95\% to 5\%. 
A far-field set consisting of about 15 hours data, and a common set consisting of about 30 hours data, are used to evaluate the performance.
\
Acoustic features are 80-dimensional energy-based log-mel filter-banks (FBK), computed on a window of 25ms with 10ms shift.
A low frame rate (LFR) is made by stacking consecutive frames into a size 7 context window (3+1+3) and then down-sampling the input frame rate to 60ms.
Acoustic modeling units are Chinese characters, which are 4233 and 9000 for AISHELL-1 and the 20,000-hour tasks respectively.
For the E2E system, all models are trained to output characters directly, without using any external LM.
\

All E2E experiments are conducted with the OpenNMT~\cite{klein2017opennmt} toolkit. We adopt the LazyAdamOptimizer with $\beta_1=0.9$, $\beta_2=0.998$, and a \emph{noam\_decay\_v2} learning rate strategy with $d=512, warmup\_n=8000$, and $k=1$~\cite{vaswani2017attention}. 
Label smoothing and dropout regularization of 0.1 are included to prevent over-fitting.

\subsection{AISHELL-1 Task}
\label{aishell1}

We first evaluate the performance on AISHELL-1.
For all system,  we set $N=6, M=3, K=0$.
The basic sub-layer output dimension, denoted $d_{basic}$, and feed-forward sub-layer $d_{ffn}$,  are set to 512 and 2048 respectively.
SpecAugment~\cite{park2019specaugment} is employed to augment the dataset.

From Table~\ref{tab:AISHLL_res1}, we see that incorporating SAN-M in a basic sub-layer obtains the best performance, compared to self-attention and DFSMN memory block.
Specially, SAN-M achieves 11.8\% relative improvement over self-attention.
From the results, it is clear that incorporating DFSMN memory blocks can boost the performance of self-attention.

We also compared the proposed SAN-M with other the popular systems in Table~\ref{tab:AISHLL_state_of_art}.
The ``LM'' column denote whether an external LM is added when decoding.
\
TDNN-LFMMI is a popular baseline reported by the dataset releaser~\cite{bu2017aishell}.
SA-T was proposed to replace the RNN with self-attention in RNN-T to obtain a performance improvements~\cite{tian2019self}.
\
LAS extended the attention-based model with an LM when decoding~\cite{shan2019component}.
\
Shigeki $et~al.$~\cite{karita2019comparative} proposed jointly training CTC and attention-based models to achieve state-of-the-art performance. Yet the proposed SAN-m system obtained slightly better performance even without using an external LM (and being more elegant).

\subsection{20,000-hour Tasks}
\label{sec:20000}

We extend our experiments to evaluate on the 20,000-hour dataset. The configuration of different systems and their results are shown in Table~\ref{tab:fly2w_res}.
For the CTC-based systems~\cite{zhang2019investigation}, we trained two DFSMN-CTC-sMBR systems with 10 and 20 DFSMN-layers, denoted CTC1 and CTC2 respectively.
$d_{basic}$ and $d_{ffn}$  are the same as described in Section~\ref{aishell1}.

Let us first compare EXP1 and EXP2.
The $Common~Set$ mainly contains near-field short duration records, and the DFSMN memory block shows comparable performance with self-attention on this task.
$Far$-$field$, which mainly contains long-duration records, highlights the superiority of self-attention at long-distance modeling.

Now comparing EXP~1 and EXP~3, we see performance improves in the system with fewer parameters, in accord with Section~\ref{aishell1}. 
This confirms that self-attention and DFSMN memory blocks are complementary, and SAN-M is able to effectively combine their strengths.
When we further explore configurations, we find that `thinner' and `deeper' structures achieve more performance improvements, as shown in EXP~4 and~5.
Compared to the EXP1 baseline, EXP5 obtains 15.3\% and 17\% relative improvements on $Common~Set$ and $Far$-$field$ tasks respectively, yet only increases the model size by less than 7\%.

\section{Conclusions}
\label{sec:conclusion}
\
In this work, we proposed memory equipped self-attention (SAN-M) to combine the strength of self-attention and DFSMN memory blocks for end-to-end speech recognition.
Our theoretical analysis and empirical comparisons concur in demonstrating the complementarity of the techniques.
This is confirmed by extensive experiments on two Mandarin ASR tasks.
\
On the AISHELL-1 task, SAN-M obtains a 11.8\% relative improvement and matches other state-of-the-art systems yet does not require an external LM.
\
Meanwhile on a 20,000-hours Mandarin ASR task, SAN-M outperforms the self-attention based Transformer baseline by over 10\%.

\vfill\pagebreak

\bibliographystyle{IEEEtran}

\bibliography{mybib}


\end{document}